\documentclass[journal]{IEEEtran}
\usepackage{xcolor,soul,framed}

\colorlet{shadecolor}{yellow}
\usepackage[pdftex]{graphicx}
\graphicspath{{../pdf/}{../jpeg/}}
\DeclareGraphicsExtensions{.pdf,.jpeg,.png}
\usepackage{graphicx}
\usepackage{caption}
\usepackage[export]{adjustbox}
\usepackage{wrapfig}
\usepackage[skip=2pt,font=footnotesize]{caption}
\usepackage{hyperref}
\hypersetup{
    colorlinks=true,
    linkcolor=blue,
    filecolor=blue,      
    urlcolor=black,
    citecolor =blue,
    pdftitle={Design of Perfect anomalous reflector},
    pdfpagemode=FullScreen
    }

\usepackage{subfig}
\usepackage[cmex10]{amsmath}
\usepackage{array}
\usepackage{mdwmath}
\usepackage{mdwtab}
\usepackage{eqparbox}
\usepackage{url}
\usepackage{textcomp}
\usepackage{gensymb}
\usepackage{multirow}
\usepackage{multicol}
\usepackage{stfloats}
\usepackage{mathrsfs}
\usepackage[noadjust]{cite}
\usepackage[ragged]{sidecap}
\usepackage{bm}% bold math
\usepackage{amssymb}
\usepackage{balance}
\usepackage{diagbox}

\newcommand{\bh}{\mathbf{h}}

\begin{document}

%\title{Efficient Multiport Coupled Scattering Synthesis of Passive-%Loaded Non-Local RISs}
\title{Efficient Scattering Synthesis for Beyond-Diagonal Non-Local RISs Coupled with Passive Load Networks}

\author{
%Sravan~Kumar~Reddy~Vuyyuru,~\IEEEmembership{Member,~IEEE,}
%Francisco~S.~Cuesta,
%Viktar~S.~Asadchy,~\IEEEmembership{Senior Member,~IEEE,}\\
%Sergei~A.~Tretyakov,~\IEEEmembership{Life Fellow,~IEEE,}
%and Do-Hoon~Kwon,~\IEEEmembership{Senior Member,~IEEE}
Sravan~Kumar~Reddy~Vuyyuru,
Francisco~S.~Cuesta,
Viktar~S.~Asadchy,
Sergei~A.~Tretyakov,
and Do-Hoon~Kwon

\thanks{This work was supported in part by the Research Council of Finland within the RCF-DoD Future Information Architecture for IoT initiative, grant no. 365679, Research Council of Finland grant no. 371367, and the U.S. National Science Foundation under grant ECCS-2448496.}
\thanks{S.~K.~R. Vuyyuru, F.~S.~Cuesta, V.~S.~Asadchy, and S.~A.~Tretyakov are with the Department of Electronics and Nanoengineering, School of Electrical Engineering, Aalto University, 02150 Espoo, Finland (e-mail: sravan.vuyyuru@aalto.fi, francisco.cuesta@aalto.fi, viktar.asadchy@aalto.fi, sergei.tretyakov@aalto.fi).}
\thanks{D.-H.~Kwon is with the Department of Electrical and Computer Engineering, University of Massachusetts Amherst, Amherst, MA 01003, USA (e-mail: dhkwon@umass.edu).}
}

\maketitle
\begin{abstract}
Realizing advanced functionalities with high efficiencies via reconfigurable intelligent surfaces (RISs) and reflectarrays requires configurations with strong electromagnetic non-local responses. The traditional approach to achieving strong non-locality has relied on modeling and synthesizing RISs with diagonal load impedance matrices composed of highly dense subwavelength structuring of arrays. In such designs, non-locality is not directly tunable, thereby limiting design flexibility and operational efficiency. This work proposes a rigorous co-simulation-based design and optimization framework for beyond-diagonal RISs with directly controllable non-locality. The co-simulation approach is based on non-local load and coupling networks, integrating electromagnetic antenna characterization with circuit-level modeling of cascaded load networks. The method benefits from additional degrees of freedom by generalizing the conventional diagonal load impedance matrix to a non-diagonal form through a non-local coupling network model. Wide-angle anomalous reflectors based on finite linear and infinite periodic arrays are designed and numerically validated, demonstrating that the proposed non-local loads embedded in realistic cascaded load networks with associated circuitry achieve significantly higher reflection efficiencies than diagonal load matrices at the given element density. Alternatively, for a fixed efficiency target, the required element density can be significantly reduced for efficient synthesis of beyond-diagonal RIS without compromising the performance of wave manipulations.
\end{abstract}


\IEEEpeerreviewmaketitle

%=== SECTION I: Introduction =================================================

\section{Introduction}\label{sec:Intro}

\IEEEPARstart{T}o address the ever-growing demand for next-generation wireless communication systems, reconfigurable intelligent surfaces (RISs) have emerged as a pivotal technology for optimizing and structuring smart radio propagation environments~\cite{Smart_Radio_Environments}. 
%The primary objective is to advance toward robust and highly reconfigurable wireless systems capable of operating in unconstrained and dynamically varying environments, thereby surpassing the limitations of current technologies. 
An RIS is an energy-efficient scattering structure equipped with tunable load components, each independently configured to dynamically and intelligently manipulate incoming electromagnetic (EM) waves to realize desired reflection or transmission properties. One of the primary functionalities of RISs is redirecting an impinging EM wave toward a predefined, arbitrary non-specular direction via anomalous reflection. In the last decade, extensive research in high-efficiency, wide-angle anomalous reflection through EM wavefront engineering has focused on modeling and designing metasurfaces~\cite{LiuKwonSATretyakov,Elefth-extreme,popov2021non,Yongming24,vuyyuru25TAP,vuyyuru25OJAP,vuyyuru23AWPL,Yongming24superdirective,li2025all,movahediqomi2025simultaneous,kwon2018lossless,diaz2017generalized,Elefth_framework,MacroscopicARM2021,almunif2025network}, metagratings~\cite{MG_YounesAlu,popov2019Feb,analytical_MG_Epstein} and metacrystals~\cite{asgari2024multifunctional} for diverse applications. Recent research is moving towards synthesizing devices for integrating multiple and reconfigurable advanced functionalities through controllable EM characteristics, such as multi-angle reflections, scanning leaky-wave antennas, and integrated sensing~\cite{popov2021non,movahediqomi2025simultaneous}.

Conventional designs of reflectarrays and reflecting metasurfaces rely on the phased-array principle and the locally periodic approximation (LPA), where a tunable reactive load impedance controls the \emph{local} reflection phase. However, phase-gradient reflectors exhibit diminished efficiencies for wide-deflection tilts, i.e., when the incidence and reflection angles deviate significantly from the conventional law of reflection. This degradation is primarily due to the wave impedance mismatch between incident and reflected waves, causing strong parasitic reflections in undesired directions~\cite{LiuKwonSATretyakov}. This limitation necessitates \emph{non-local} designs of RISs, where the performance of wavefront control of reflected waves is enhanced by properly optimizing coupling between scatterers. For these reasons, in the antenna and electromagnetics community, non-local designs of reconfigurable aperiodically loaded metasurfaces have become prevalent~\cite{diaz2017generalized,vuyyuru25TAP,kwon2018lossless,popov2021non,Elefth-extreme,Elefth_framework,Yongming24,Yongming24superdirective,vuyyuru23AWPL,vuyyuru25OJAP,movahediqomi2025simultaneous}. For instance, recent research demonstrates that non-local designs incorporating spatially varying subwavelength-sized meta-atoms can achieve nearly 100\%-efficient scanning anomalous reflectors, with array scatterers modeled as impedance wires or thin strips~\cite{Yongming24} and realistic patches~\cite{vuyyuru23AWPL,vuyyuru25TAP,vuyyuru25OJAP}. Furthermore, optimizing evanescent fields in finite-size arrays by inducing oscillatory current distribution in subwavelength-engineered non-local RISs enables a superdirective behavior that surpasses the conventional 100\% efficiency limit of infinite arrays. This has been demonstrated both in idealized thin-wire models~\cite{Yongming24superdirective,li2025all} and in practical patch-array implementations~\cite{movahediqomi2025simultaneous}. Such superdirectivity and advanced functionalities demand additional degrees of freedom for optimization of coupling between array elements, which have been conventionally achieved through deep subwavelength discretization, i.e., by increasing the number of scatterers per wavelength. However, densifying the array by reducing the element spacing introduces practical challenges, including complex layouts, fabrication constraints, scalability issues due to increased losses, strong mutual coupling, and increased calibration complexity. On the other hand, metagratings avoid the need for a subwavelength spacing but restrict continuous scanning, as their fixed periodicity enables scanning only at specific discrete angles determined by the supported Bloch harmonics for a given set of incidence and reflection angles~\cite{MG_YounesAlu,popov2019Feb,analytical_MG_Epstein}.

Interestingly, recent advances in the communications community have also recognized the necessity for non-local RIS designs, introducing the concept of Beyond-Diagonal RISs (BD-RISs)~\cite{shen2021modeling,Li2023Beyond,Li2024RIS2,li2025tutorial,Li2024mutualcoupling,Philipp2025PC,Tapie2025BDRISPrototype,ming2025hybrid}. In traditional communication-theory models of RIS structures, array elements are assumed to behave as individual, independently controlled scatterers, a configuration widely used in existing studies~\cite{Fotock24,Smart_Radio_Environments}. In contrast, proposed BD-RIS designs interconnect array elements through a general reconfigurable impedance load network defined by a non-diagonal load matrix, introducing controlled cross-coupling using inter-element connections~\cite{shen2021modeling,Li2023Beyond,Li2024RIS2,li2025tutorial}. This controllable mutual coupling between RIS elements can be exploited to further enhance the performance of the RIS. However, these formulations typically assume that RIS elements are electromagnetically decoupled, i.e., treated as isolated elements while neglecting mutual interactions through near fields between the array elements. In other words, most existing BD-RIS models do not account for physical EM coupling of the array elements, modeling the RIS response with an oversimplified matrix of phase shifts introduced by idealized reflecting elements. The authors in~\cite{Li2024mutualcoupling,Philipp2025PC,Tapie2025BDRISPrototype,ming2025hybrid} proposed coupling-aware models for BD-RISs that capture mutual coupling at the system level. However, these models rely on pre-characterization of the complete communication channel and, therefore, are limited to static or quasi-static communication environments. For instance, every transmitter or receiver movement requires load recalibration, whereas the proposed antenna-based BD-RIS model enables receiver/transmitter-position-agnostic synthesis. Meanwhile, in the EM community, engineered non-local metasurfaces with spatially varying subwavelength-spaced elements~\cite{Yongming24,Yongming24superdirective,vuyyuru23AWPL,vuyyuru25TAP,vuyyuru25OJAP,movahediqomi2025simultaneous} have focused on controlling inter-element mutual coupling by globally synthesizing the diagonal load impedances to enhance wave-matter interactions, without the need for a non-diagonal load matrix.

\begin{figure*}[t]
\centering
\includegraphics[width=7.1in,height = 2.5in]{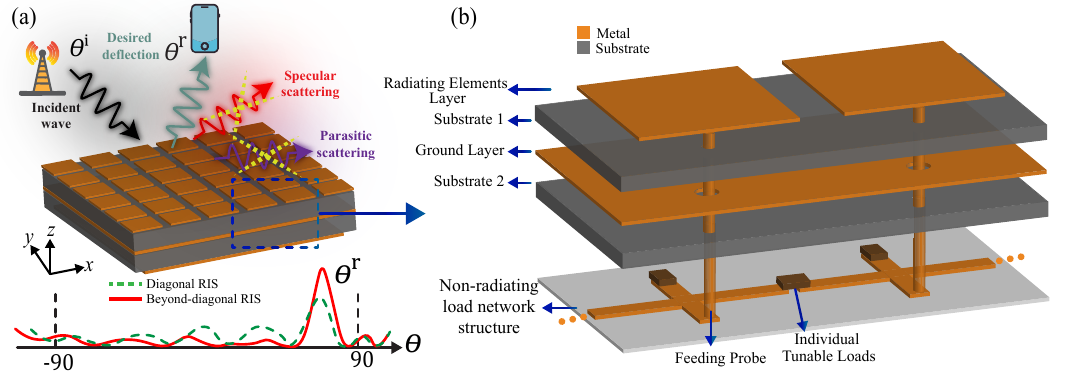}
\caption{(a) 3D schematic of the multilayer BD-RIS structure composed of loaded radiating elements, illustrating plane-wave incidence at angle $\theta^{\rm i}$ and anomalous deflection toward the desired angle $\theta^{\rm r}$. The lowermost plot presents the radiation characteristics, indicating improved sidelobe suppression and higher reflected-power efficiency relative to the diagonal RIS configuration. (b) A magnified, exploded view of a two-element multilayer unit cell stack-up highlighting the transmission-line-based interconnection and individually tunable loads embedded within the non-radiating tunable layer.}
\label{fig:BD_RIS}
\end{figure*}

%From the foregoing modeling discussion, it becomes evident that the metasurface EM community has developed scattering-synthesis frameworks that account for inter-element coupling through physical near-field interactions above the ground plane. However, such designs are constrained by practical and physical challenges associated with array densification. Conversely, the communications community models coupling abstractly by optimizing beyond-diagonal coupling load network connections, mostly neglecting the underlying physical interactions within the array.

In this paper, to overcome these limitations, we propose a strategy that fully synthesizes EM coupling of BD-RIS elements by introducing an additional non-radiating, tunable network layer beneath the ground plane, interconnecting the element ports, as illustrated in Fig.~\ref{fig:BD_RIS}. This approach is particularly relevant to RIS arrays with larger inter-element spacing (e.g., typically half-wavelength), as it bypasses the subwavelength structuring requirement for fine-tuning non-local responses. The proposed strategy fully accounts for all field interactions above the ground plane and the guided-wave coupling using a distributed-circuit load network beneath the ground plane, enabling ideal and even superdirective performance for conventional half-wavelength-spacing RIS arrays over all scan angles. The non-radiating tunable load network introduces a virtual array with advanced controllable ports, raising the degrees of freedom for efficiently controlling mutual coupling among the array elements. Moreover, positioning the coupling network behind the ground plane isolates it from environmental perturbations, thereby improving robustness, scalability, and design stability.

\begin{figure}[t]
\centering
\includegraphics[width=3.45in,height = 1.5in]{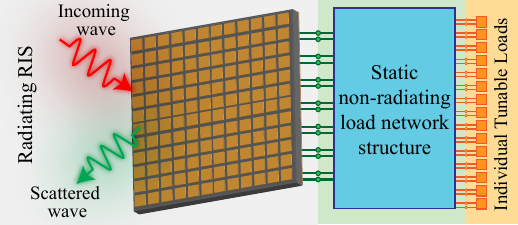}
\caption{Conceptual cascaded multiport network representation of the proposed methodology. The RIS element ports are terminated by a composite load network, decomposed into a static non-radiating structure and individually tunable lumped loads, which emulates more controllable EM degrees of freedom beyond the number of physical RIS elements.}
\label{fig:cascaded}
\end{figure}

In previous generations, classical scattering arrays utilized interconnections between element ports behind the ground plane to realize non-local coupling, and they are popularly known as retrodirective arrays~\cite{Miyamoto2002retro,Chung1998retro}, or retroreflectors, that operate for arbitrary angles of incidence. These retroreflectors redirect the beams towards the source through phase-conjugating connections between spatially separated elements, to transfer phase and amplitude information via non-radiating feed networks. The proposed approach represents a broad generalization of this concept, pairing the radiating array with additional coupling transmission lines and lumped impedance loads positioned behind the ground plane to synthesize both diagonal and off-diagonal load terminations using a generalized multiport coupled network formulation, as portrayed in Fig.~\ref{fig:cascaded}. Designing fully EM-compatible load networks and integrating them with physical arrays will bridge the gap between communications and electromagnetics communities in the analysis and design of scattering arrays paired with non-diagonal load networks. Introducing non-radiating coupling networks requires their design and realization as distributed circuits, which are not needed in subwavelength structured arrays. Nevertheless, the efficiency, accuracy, and low cost of their design and implementation, especially in printed circuit technologies, makes the proposed architecture a preferred choice compared with dense arrays.

This research paper is organized as follows. Section~\ref{sec:methodology} introduces the proposed methodology of the cascaded multiport network. Section~\ref{sec:results} presents computation results for optimized aperiodically loaded patch arrays with two arbitrary array configurations realizing anomalous deflection without subwavelength structuring. The general summary and concluding remarks are provided in Section~\ref{sec:conclusion}.

\section{Formulation of Multiport Coupled Network Methodology}\label{sec:methodology}

Previous works proposed a multi-objective nonlinear algebraic array antenna scattering synthesis framework to optimize RIS scattering characteristics by accurately predicting induced current distributions under passive-loaded lumped elements~\cite{vuyyuru23AWPL,vuyyuru25OJAP}. Preliminary full-wave EM simulations are performed in transmission (Tx) and reflection (Rx) configurations to extract the scattered E-field under open-circuited load conditions (i.e., structural scattering contribution), the array $Z$-parameter matrix comprising both self (diagonal) and mutual (symmetrical off-diagonal) impedances, open-circuit terminal voltages, and vector effective heights. They form the basis for scattering synthesis and radiation pattern optimization. For instance, these parameters are used as input to an algebraic global optimization to maximize the radiated power efficiency in the desired direction. However, subwavelength-spaced structuring is required to achieve a high anomalous scattering efficiency, leading to practical challenges associated with array densification. In this section, we develop a multiport network model to design a non-local beamforming metasurface by synthesizing the scattering response of the structure terminated with an arbitrary load network.  

\subsection{Core Idea of the Proposed Methodology}\label{subsec:core_idea}

Although the proposed methodology can be applied to arbitrary reflectarrays, for clarity, we present it for a specific example of a microstrip patch array whose elements are printed on a grounded dielectric substrate of thickness $h$. Beneath the ground plane, an additional dielectric substrate is introduced, followed by a tunable metallic control layer located beneath this substrate, interconnecting the feeding ports of the elements, forming a vertically stacked multilayer structure, as shown in Fig.~\ref{fig:BD_RIS}. Metallic vias connect the radiating elements to the tunable layer. 

The general concept of the proposed BD-RIS configuration is illustrated in Figs.~\ref{fig:BD_RIS} and \ref{fig:cascaded}. This and similar scattering systems can be directly analyzed and optimized by considering connections to all individual tunable loads as conventional antenna ports. Following the same method as developed earlier in \cite{vuyyuru25OJAP}, for each of these ports, one can calculate the angle-dependent vector effective height. Likewise, one can calculate the impedance matrix of the whole structure connected to the tunable loads. Importantly, in this scenario, the geometrical layout of the radiating structure can be extremely general. It can be an array of an arbitrary number of radiating unit cells or even a single radiating body that is connected to our individual tunable loads at several different points. Basically, this device is a multi-mode antenna. Due to an increased number of modes combined with guided-wave coupling through the load network, we expect to achieve high performance for any scan angle without using subwavelength-sized unit elements. A specific example in Fig.~\ref{fig:cascaded} illustrates the case when the radiating structure is a conventional array of half-wavelength-sized unit cells, which has the advantage of resonant behavior of each radiating element. However, incorporating a load network located beneath the ground plane makes full-wave EM characterization of the combined antenna array and load network computationally expensive and time-consuming, particularly for electrically large or multiport configurations. On the other hand, EM simulation accurately accounts for all inter-port coupling.

To mitigate this computational cost, the main idea is to adopt a hybrid co-simulation workflow by characterizing the radiating array structure via full-wave simulations (e.g., using CST Studio Suite), while the distributed-circuit feed (load) network beneath the ground plane is modeled separately at the circuit level (e.g., using Keysight Advanced Design System -- ADS) to extract its scattering parameters, as illustrated in Fig.~\ref{fig:cascaded}. Provided that the feed (load) network is non-radiating (i.e., no radiation leakage or unwanted EM coupling) and accurately characterized by circuit simulation, this approach is substantially more efficient than brute-force EM characterization of the entire structure. The main advantage of cascaded network treatment (a radiating array + a non-radiating network) is that EM characterization of the array captures all EM field interactions above the ground plane, while circuit-level modeling of the feed (load) network accounts for guided-wave coupling beneath it accurately within an EM-consistent design framework. This decoupled EM–circuit co-simulation framework preserves physical accuracy while reducing computational complexity, enabling practical large-scale non-diagonal RIS design and efficient scattering optimization.

The following subsections present the cascaded network representation of the BD-RIS coupled with a feed network terminated in tunable impedance loads. Two equivalent array-load cascade networks are analyzed. The total scattered field is decomposed into structural and Tx-mode scattering contributions, expressed in terms of network parameters, vector effective heights, and array received voltages.
%derive the input impedance matrix, open-circuit voltages, and vector effective heights of the feed network using a multiport scattering system formulation~\cite{Kuznetsov24,Nie2014}. These parameters will enable the prediction and synthesis of the scattering response of non-diagonal RIS terminated with arbitrary loads~\cite{Kim25Eff,vuyyuru25OJAP}.

\subsection{Scattering Control Configuration}\label{subsec:Scattering}

\begin{figure}[tb]
\centering
\includegraphics[width=3.48in,height = 1.8in]{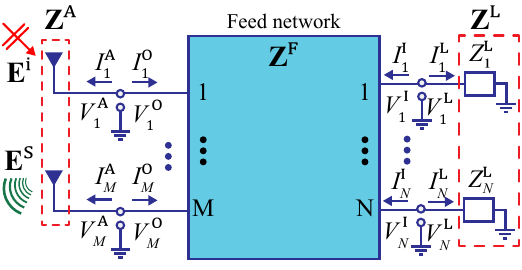}
\caption{Illustration of a cascaded multiport network formed by interconnecting an array antenna structure with $M$ ports with a feed network terminated in $N$ isolated loads. The feed network is characterized by an $(M+N)\times(M+N)$ $Z$-parameter matrix $\mathbf{Z}^{\rm F}$.}
\label{fig3:Scattering}
\end{figure}

Figure~\ref{fig3:Scattering} illustrates the block-diagram representation of the cascaded multiport network, composed of physical array elements on the left and isolated (diagonal) loads on the right, interconnected through a feed network. The array is illuminated by an incident plane wave with an $E$-field $\mathbf{E}^{\rm i}$ (under the $e^{j\omega t}$ time convention), and the loaded array generates the scattered $E$-field $\mathbf{E}^{\rm s}$. The array structure, comprising $M$ radiating elements, is modeled as an $M$-port network characterized by the $Z$-parameter matrix $\mathbf{Z}^{\rm A} \in \mathbb{C}^{M \times M}$. In a Tx scenario, the port voltages and currents satisfy the impedance relation, \mbox{$\mathbf{V}^{\rm A} = \mathbf{Z}^{\rm A}\mathbf{I}^{\rm A}$}, where \mbox{$\mathbf{V}^{\rm A}=[V^{\rm A}_1,\cdots,V^{\rm A}_M]^{\rm T}$} and \mbox{$\mathbf{I}^{\rm A}=[I^{\rm A}_1,\cdots,I^{\rm A}_M]^{\rm T}$}. As illustrated in Fig.~\ref{fig:cascaded}, the non-radiating network can be viewed as two cascaded sub-networks: an invariant bridge multiport feed network and a tunable, reactively-terminated load network~\cite{Philipp2025PC}. The tunable load network is modeled as an $N$-port linear network characterized by the diagonal $Z$-matrix $\mathbf{Z}^{\rm L} \in \mathbb{C}^{N \times N}$. The corresponding load voltage and current vectors are \mbox{$\mathbf{V}^{\rm L}=[V^{\rm L}_1,\cdots,V^{\rm L}_M]^{\rm T}$} and \mbox{$\mathbf{I}^{\rm L}=[I^{\rm L}_1,\cdots,I^{\rm L}_M]^{\rm T}$}, satisfying \mbox{$\mathbf{V}^{\rm L} = \mathbf{Z}^{\rm L}\mathbf{I}^{\rm L}$}. 

In contrast, the bridge multiport feed network is introduced as a static interconnection between the $N$ load ports and the $M$ antenna ports. Let us denote by $\mathbf{V}^{\rm I}, \mathbf{I}^{\rm I} \in \mathbb{C}^{N}$ the input-port variables for the feed network (the load side in an Rx scenario) and by $\mathbf{V}^{\rm O}, \mathbf{I}^{\rm O} \in \mathbb{C}^{M}$ the output-port variables for the feed network (the antenna side in an Rx scenario) with \mbox{$\mathbf{V}^{\rm O}=[V^{\rm O}_1,\cdots,V^{\rm O}_M]^{\rm T}$}, \mbox{$\mathbf{I}^{\rm O}=[I^{\rm O}_1,\cdots,I^{\rm O}_M]^{\rm T}$}, \mbox{$\mathbf{V}^{\rm I}=[V^{\rm I}_1,\cdots,V^{\rm I}_N]^{\rm T}$}, and \mbox{$\mathbf{I}^{\rm I}=[I^{\rm I}_1,\cdots,I^{\rm I}_N]^{\rm T}$}. Port interconnection interfaces enforce continuity relations $\mathbf{V}^{\rm A} = \mathbf{V}^{\rm O}$, $\mathbf{I}^{\rm A} = -\mathbf{I}^{\rm O}$ on the output interface and $\mathbf{V}^{\rm I} = \mathbf{V}^{\rm L}$, $\mathbf{I}^{\rm I} = -\mathbf{I}^{\rm L}$ on the input interface of the feed network. The negative signs arise from opposite current reference directions at each port interface. The resulting cascaded mulitport feed network is characterized by the overall \mbox{Z-parameter} matrix $\mathbf{Z}^{\rm F} \in \mathbb{C}^{(M+N)\times(M+N)}$, written in a block matrix form as
\begin{equation}
\mathbf{Z}^{\rm F}=\begin{bmatrix}
\mathbf{Z}^{\rm OO} & \mathbf{Z}^{\rm OI} \\
\mathbf{Z}^{\rm IO} & \mathbf{Z}^{\rm II}
\end{bmatrix}.\label{eq:ZF}
\end{equation}
The $Z$-parameter matrix $\mathbf{Z}^{\rm F}$ can be partitioned into four submatrices associated with the input (loads) and output (patchs) sides of the matching (feed) network. Each of the four submatrices represents the voltage-current relation between ports: the $M \times M$ submatrix $\mathbf{Z}^{\rm OO}$ relates the $M$ element ports; the $N \times N$ submatrix $\mathbf{Z}^{\rm II}$ relates the $N$ discrete load ports; and the $M \times N$ $\mathbf{Z}^{\rm OI}$ and $N \times M$ $\mathbf{Z}^{\rm IO}$ submatrices relate the antenna and load ports. The block-matrix voltage-current relation is
\begin{equation}\begin{bmatrix}
\mathbf{V}^{\rm O} \\ \mathbf{V}^{\rm I}
\end{bmatrix} =
\mathbf Z^{\rm F}
\begin{bmatrix}
\mathbf{I}^{\rm O} \\ \mathbf{I}^{\rm I}
\end{bmatrix} =
\begin{bmatrix}
\mathbf{Z}^{\rm OO} & \mathbf{Z}^{\rm OI} \\
\mathbf{Z}^{\rm IO} & \mathbf{Z}^{\rm II}
\end{bmatrix}
\begin{bmatrix}
\mathbf{I}^{\rm O} \\ \mathbf{I}^{\rm I}
\end{bmatrix}.
\label{eq:ZF_expansion}\end{equation}

The cascade of three networks in Fig.~\ref{fig3:Scattering} can be interpreted as an equivalent standard source-load cascade of two networks. There are two different interpretations. The feed network-loads cascade can be viewed as an equivalent $M$-port load network. Alternatively, the array-feed network cascade can be interpreted as an equivalent $N$-element array.

\subsection{An M-Element Antenna Array Terminated in an Equivalent Full Load Network}\label{subsec:MEle-MLoadPort}

\begin{figure}[h]
\centering
\includegraphics[width=3.45in,height = 1.7in]{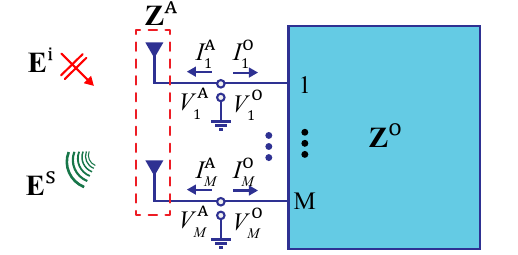}
\caption{Illustration of the composite load network (the feed network and isolated loads) as seen from the antenna structure. In this scenario the $M$-element antenna array are connected to a $M$-port network with equivalent full load-matrix $\mathbf{Z}^{\rm O}$. This configuration is similar to the one used to optimize diagonal RISs.}
\label{fig4:MEle-MLoadPort}
\end{figure}

In this viewpoint, the original cascaded antenna–feed network–load system is reduced to an equivalent representation, where an $M$-element antenna array is terminated by an $M$-port load network with an impedance matrix, $\mathbf{Z}^{\rm O}$, as depicted in Fig.~\ref{fig4:MEle-MLoadPort}. The formulation follows the same formal framework adopted in the conventional diagonal load configuration. First, a preliminary numerical full-wave EM simulation of the $M$-element array is performed to extract the required EM parameters: the input impedance matrix $\mathbf{Z}^{\rm A}$, the open-circuit voltage $\mathbf{V}^{\rm A}_\text{oc}$, the vector effective height $\bar{\bh}^\text{A}(\theta^{\rm r})$, and the structural scattering field $\mathbf{E}^{\rm s}(\mathbf{I}^\mathrm{A}=0)$. The isolated-load diagonal impedance matrix $\mathbf{Z}^{\rm L}$ is now generalized to a full load-network matrix $\mathbf{Z}^{\rm O}$, which is the impedance matrix looking from the array element ports into the feed network terminated by $\mathbf{Z}^\text{L}$, see Fig.~\ref{fig3:Scattering}. To derive $\mathbf{Z}^{\rm O}$, we start from the general composite port impedance $Z$-matrix of the feed network in~\eqref{eq:ZF_expansion} written as
\begin{equation}\begin{bmatrix}
\mathbf{V}^{\rm O} \\ \mathbf{V}^{\rm L}
\end{bmatrix} =
\begin{bmatrix}
\mathbf{Z}^{\rm OO} & \mathbf{Z}^{\rm OI} \\
\mathbf{Z}^{\rm IO} & \mathbf{Z}^{\rm II}
\end{bmatrix}
\begin{bmatrix}
\mathbf{I}^{\rm O} \\ -\mathbf{I}^{\rm L}
\end{bmatrix}.
\label{eq:ZO_expansion}\end{equation}
Expanding the second equation in Eq.~\eqref{eq:ZO_expansion} and using $\mathbf{V}^\text{L}=\mathbf{Z}^\text{L} \mathbf{I}^\text{L}$ yields
\begin{equation}
    \mathbf{I}^{\rm L} = 
    (\mathbf{Z}^{\rm L}+\mathbf{Z}^{\rm II})^{-1}
    \mathbf{Z}^{\rm IO}\mathbf{I}^{\rm O}.
    \label{eq:ZO_IL}
\end{equation}
Substituting~\eqref{eq:ZO_IL} into the first equation of Eq.~\eqref{eq:ZO_expansion} yields
\begin{equation}
    \mathbf{V}^{\rm O} = 
    \left[\mathbf{Z}^{\rm OO}-\mathbf{Z}^{\rm OI}(\mathbf{Z}^{\rm L}+\mathbf{Z}^{\rm II})^{-1}
    \mathbf{Z}^{\rm IO}\right]\mathbf{I}^{\rm O}.
    \label{eq:ZO_VO}
\end{equation}
Hence, the $M$-port load impedance $Z$-matrix is identified from~\eqref{eq:ZO_VO} as
\begin{equation}
    \mathbf{Z}^{\rm O} =\mathbf{Z}^{\rm OO}-
    \mathbf{Z}^{\rm OI}(\mathbf{Z}^{\rm L}+\mathbf{Z}^{\rm II})^{-1}\mathbf{Z}^{\rm IO}.
    \label{eq:ZO}
\end{equation}
This result represents the equivalent $M$-port load network seen at the antenna ports.

\subsection{An equivalent $N$-port coupled array terminated in $N$ Isolated Loads}\label{subsec:NEle-NLoadPort}

\begin{figure}[h]
\centering
\includegraphics[width=3.48in,height = 2in]{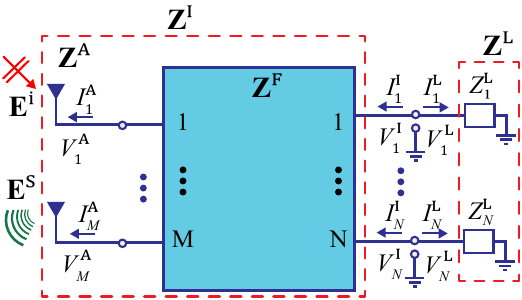}
\caption{Schematic representing the interaction of the $N$ isolated loads with the $M$ antenna elements through the feed network. From the loads perspective, the antennas and the feed network are seen as an equivalent $N$-element array with an equivalent impedance matrix $\mathbf{Z}^{\rm I}$. This approach is also useful to determine the vector effective heights $\bar{\bh}^\text{I}$ for a BD-RIS.}
\label{fig5:NEle-NLoadPort}
\end{figure}

Figure~\ref{fig5:NEle-NLoadPort} depicts an equivalent $N$-element receiving array configuration modeled as a linear multiport network characterized by its impedance matrix, $\mathbf{Z}^{\rm I}$, with each port terminated in an isolated load represented by the diagonal matrix, $\mathbf{Z}^{\rm L}$. For BD-RIS, a co-simulation workflow based on a cascaded multiport network formed by interconnecting the $M$-element physical array with the $M$ output ports of the feed network makes an equivalent $N$-element array. A full-wave numerical EM simulation of the $M$-element physical array is conducted to extract the required EM parameters. The non-radiating feed network is characterized at the circuit level.

The network parameters of the array-feed network cascade can be derived as follows. First, Eq.~\eqref{eq:ZF_expansion} is written in terms of $\mathbf{V}^\text{A}$ and $\mathbf{I}^\text{A}$ as
\begin{equation}\begin{bmatrix}
\mathbf{V}^{\rm A} \\ \mathbf{V}^{\rm I}
\end{bmatrix} =
\begin{bmatrix}
\mathbf{Z}^{\rm OO} & \mathbf{Z}^{\rm OI} \\
\mathbf{Z}^{\rm IO} & \mathbf{Z}^{\rm II}
\end{bmatrix}
\begin{bmatrix}
-\mathbf{I}^{\rm A} \\ \mathbf{I}^{\rm I}
\end{bmatrix}.
\label{eq:ZI_expansion}\end{equation}
Expanding the first equation in Eq.~\eqref{eq:ZI_expansion} and using $\mathbf{V}^\text{A}=\mathbf{Z}^\text{A} \mathbf{I}^\text{A}$ yields
\begin{equation}
    \mathbf{I}^{\rm A} = 
    (\mathbf{Z}^{\rm A}+\mathbf{Z}^{\rm OO})^{-1}
    \mathbf{Z}^{\rm OI}\mathbf{I}^{\rm I}.
    \label{eq:ZI_IA}
\end{equation}
Substituting Eq.~\eqref{eq:ZI_IA} into the second equation of Eq.~\eqref{eq:ZI_expansion} yields
\begin{equation}
    \mathbf{V}^{\rm I} = 
    \left[\mathbf{Z}^{\rm II}-\mathbf{Z}^{\rm IO}(\mathbf{Z}^{\rm A}+\mathbf{Z}^{\rm OO})^{-1}
    \mathbf{Z}^{\rm OI}\right]\mathbf{I}^{\rm I}.
    \label{eq:ZI_VI}
\end{equation}
The $N$-port input impedance $Z$-matrix is identified from Eq.~\eqref{eq:ZI_VI} as
\begin{equation}
    \mathbf{Z}^{\rm I} =\mathbf{Z}^{\rm II}-\mathbf{Z}^{\rm IO}(\mathbf{Z}^{\rm A}+\mathbf{Z}^{\rm OO})^{-1}
    \mathbf{Z}^{\rm OI}.
    \label{eq:ZI}
\end{equation}

For open-circuit voltages at the $N$ load ports, consider voltage source excitation of the network at the array element ports, while the load ports on the right side are left open-circuited. The source voltages are the received array element port voltages when all element ports are open-circuited. They are denoted by \mbox{$\mathbf{V}^{\rm A}_\text{oc} =[V^{\rm A}_{\text{oc}, 1}, \cdots, V^{\rm A}_{\text{oc}, M}]^{\rm T}$}, and the induced open-circuit voltages at the input (i.e., isolated load) ports are denoted by \mbox{$\mathbf{V}^{\rm I}_\text{oc} =[V^{\rm I}_{\text{oc}, 1}, \cdots, V^{\rm I}_{\text{oc}, N}]^{\rm T}$}. The currents flowing into the input and output ports are $\mathbf{I}^{\rm I}$ and $-\mathbf{I}^{\rm A}$, respectively. Under open-circuit termination of the $N$ input ports, $\mathbf{I}^{\rm I} =0$, the voltage–current relation of the feed network gives
\begin{equation}\begin{bmatrix}
\mathbf{V}^{\rm A}_\text{oc} \\ \mathbf{V}^{\rm I}_\text{oc}
\end{bmatrix} =
\mathbf Z^{\rm F}
\begin{bmatrix}
-\mathbf{I}^{\rm A} \\ \mathbf{I}^{\rm I}
\end{bmatrix} =
\begin{bmatrix}
\mathbf{Z}^{\rm OO} & \mathbf{Z}^{\rm OI} \\
\mathbf{Z}^{\rm IO} & \mathbf{Z}^{\rm II}
\end{bmatrix}
\begin{bmatrix}
-\mathbf{I}^{\rm A} \\ 0
\end{bmatrix}.
\label{eq:Voc_expansion}
\end{equation}
The $N\times 1$ column open-circuit voltage vector on the input side feed ports is acquired from Eq.~\eqref{eq:Voc_expansion} as
\begin{equation}
    \mathbf{V}^{\rm I}_\text{oc} = \mathbf{Z}^{\rm IO}
    (\mathbf{Z}^{\rm OO})^{-1}
    \mathbf{V}^{\rm A}_\text{oc}.
    \label{eq:Voc}
\end{equation}

Next, the vector effective heights referenced at the input ports of the equivalent $N$-element array should be found. The array element ports are associated with an array of element vector effective heights \mbox{$\bar{\bh}^\text{A}=[\bh_1^\text{A},\cdots,\bh_M^\text{A}]^\text{T}$}, while those of the input (load) ports are characterized by \mbox{$\bar{\bh}^\text{I}=[\bh_1^\text{I},\cdots,\bh_N^\text{I}]^\text{T}$}. The overbar indicates that the elements of $\bar{\bh}^\text{A}$ and $\bar{\bh}^\text{I}$ are vectors. The effective height of an element in an array represents the far-zone vector pattern when only the element under consideration is excited and all other elements are open-circuited~\cite{Kim25Eff}. Hence, the relation between $\mathbf{I}^\text{A}$ and $\mathbf{I}^\text{I}$ determines $\bar{\bh}^\text{I}$ in terms of $\bar{\bh}^\text{A}$. From Eq.~\eqref{eq:ZI_IA}, we find that the $n$-th column of the $M\times N$ matrix $(\mathbf{Z}^\text{A}+\mathbf{Z}^\text{OO})^{-1}\mathbf{Z}^\text{OI}$ represents the $M$ array element port currents when a unit current flows into the $n$-th input port, while all other input ports are open-circuited. Therefore, we find
\begin{equation}
    \bar{\bh}^\text{I} = 
    [(\mathbf{Z}^{\rm A}+\mathbf{Z}^{\rm OO})^{-1}
    \mathbf{Z}^{\rm OI}]^\text{T}\bar{\bh}^\text{A}.
    \label{eq:he}
\end{equation}

\begin{figure}[h]
\centering
\includegraphics[width=3.45in,height = 1.6in]{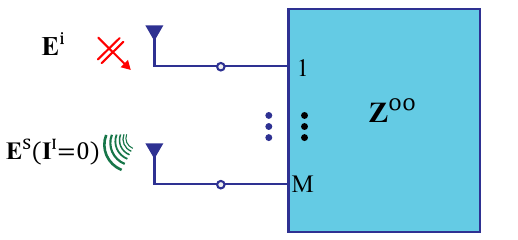}
\caption{Equivalent configuration in which the $M$-element array is terminated by a load network characterized by the impedance matrix $\mathbf{Z}^{\rm OO}$ for determining the structural scattering term for the equivalent $N$-element array.}
\label{fig6:Structural}
\end{figure}

 The input impedance matrix $\mathbf{Z}^\text{I}$, open-circuit voltages $\mathbf{V}^\text{I}_\text{oc}$, and vector effective heights of the equivalent $N$-element array $\bar{\bh}^\text{I}$ are computed using the above network derivations. The structural scattering term $\mathbf{E}^{\rm s}(\mathbf{I}^\text{I}=0)$ represents scattering by the array when the feed network is terminated in $N$ open-circuit loads, $\mathbf{Z}^\mathrm{L}=\infty$. Here, the array element port currents $\mathbf{I}^\text{A}$ are not expected to be zero. In finding $\mathbf{E}^\text{s}(\mathbf{I}^\text{I}=0)$, the feed network need not be physically modeled in numerical EM simulations. Instead, as illustrated in Fig.~\ref{fig6:Structural}, the $M$-element array is terminated by an equivalent load network characterized by the $Z$-parameter matrix $\mathbf{Z}^{\rm OO}$, which $\mathbf{Z}^\text{O}$ in Eq.~\eqref{eq:ZO} reduces to when $\mathbf{Z}^\text{L}=\infty$.

 For the scattering synthesis configuration in Fig.~\ref{fig3:Scattering}, the two treatments in Sections~\ref{subsec:MEle-MLoadPort} and \ref{subsec:NEle-NLoadPort} are equivalent. Taking non-diagonal coupling via the feed network fully into account, scattering synthesis is performed via algebraic optimization of a diagonal reactive load matrix defined in the $N$-port network, providing additional degrees of freedom without requiring sub-wavelength structuring.
 %As the equivalent $M$-port load network treatment in Section~\ref{subsec:MEle-MLoadPort} is a straightforward extension to the diagonal load optimization approach, we adopt this method in the following scattering  synthesis examples.}
In this work, we adopt the method explained in Section~\ref{subsec:MEle-MLoadPort} in the following scattering synthesis examples.

\section{Numerical Optimization and Analysis of Patch Arrays}\label{sec:results}

This section presents and analyzes the performance characteristics of the proposed BD-RIS architecture using two scattering configurations: a supercell-based infinite periodic planar reflector and a finite linear-array scatterer. The proposed structures target distinct EM design objectives. The infinite periodic reflector aims to redirect an incident propagating plane wave into another direction to achieve a high anomalous reflection efficiency. In contrast, the objective of the finite reflector under plane-wave illumination is to redirect the incoming power toward the desired direction, maximizing the scattered-field intensity in the main beam while suppressing parasitic scattering into undesired directions.

For both design configurations, a consistent set of steps are adopted, beginning with unit-cell simulations of the RIS in CST using periodic boundary conditions to determine the geometrical parameters of a self-resonant patch antenna. Specifically, the side length and feed position of a linearly polarized perfect electric conductor (PEC) square patch are tuned for a $50~\Omega$ impedance match at $28$~GHz (with a corresponding free-space wavelength $\lambda\approx10.71$~mm). Each unit cell consists of a square PEC patch printed on a lossless Rogers RO4350B substrate ($\epsilon_r$ = 3.66, $h =0.338$~mm) backed by a PEC ground plane, with the optimized geometrical parameters summarized in Table~\ref{tab:dimentions}.

\begin{table}[t]
\renewcommand{\arraystretch}{1.25}
\centering
    \caption{RIS dimensions}
    \begin{tabular}{|c|c|c|} \hline
    Parameters & Supercell array & Finite linear array \\ \hline
    Array dimension & $1.064\lambda\times0.3547\lambda$ & $10.5\lambda\times0.5\lambda$\\ \hline
    Unit cell spacing & $0.3547\lambda$ & $0.5\lambda$ \\ \hline
    Square patch dimension & $0.2471\lambda $  & $0.2385\lambda$\\ \hline
    Probe feed position & $(0,1.1093)$ mm & $(0,0.444)$ mm \\ \hline
    Total elements $(N)$ &  $3$  & $21$ \\ \hline
    Substrate height ($h$) & \multicolumn{2}{c|}{$0.338$~mm} \\ \hline
    \end{tabular}
    \label{tab:dimentions}
\end{table}

\subsection{Periodic RISs supporting three Bloch diffraction orders} \label{subsec:PeriodicRIS}

\begin{figure*}[ht]
 \centering
\includegraphics[width=1\linewidth]{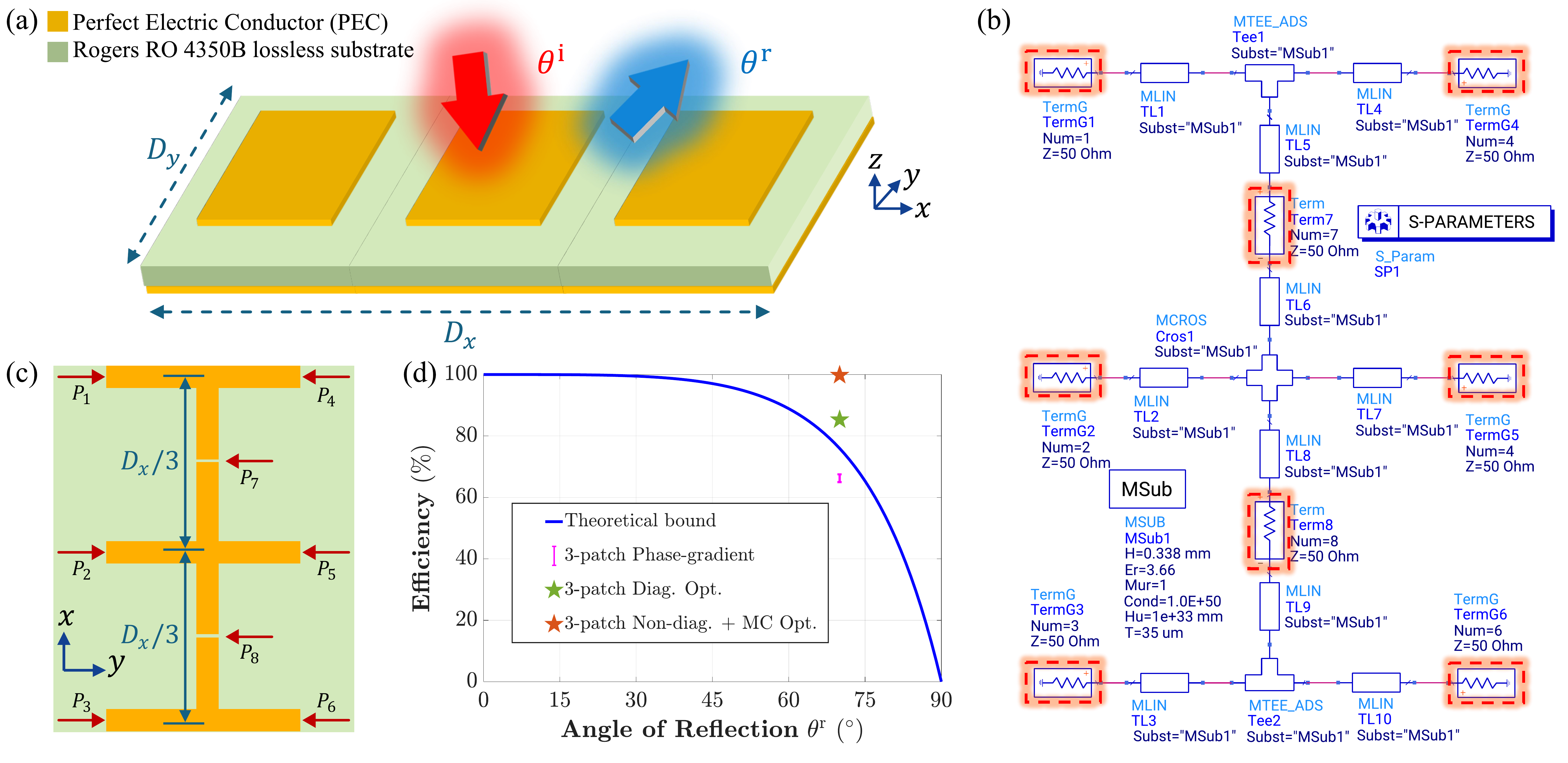}
\caption{(a) Periodic RIS with a three-patch supercell illustrating anomalous reflection from $\theta^{\mathrm{i}}$ to $\theta^{\mathrm{r}}$. (b) Schematic of the distributed ADS transmission-line network model for the 8-port feed network. (c) Generated physical layout corresponding to the equivalent ADS transmission-line model depicted in (b) with port locations. (d) Power efficiency comparison between the optimized and conventional phase-gradient designs at $\theta^{\mathrm{r}} = 70^\circ$. The blue curve represents the theoretical efficiency of the phase-gradient reflector~\cite{asadchy2017eliminating}.} 
\label{fig:RISPeriodResults}
\end{figure*}

An anomalous reflector based on an infinite periodic supercell of dimensions $D_x$ and $D_y$, along the $x$- and $y$-directions, respectively, is realized utilizing a three-patch supercell linearly aligned along the $x$-axis. They are tailored for transverse electric (TE) polarization under normal plane-wave incidence, i.e., $\theta^{\rm i}=0^\circ$, as depicted in Fig.~\ref{fig:RISPeriodResults}(a). Following the basics of the Bloch theory, the supercell period is defined as $D_x=\lambda/\sin\theta^{\rm r}$, which allows three Bloch propagating diffraction harmonics at the specular direction and predetermined deflection angles of \mbox{$\theta=\pm\theta^{\rm r}=\pm70^\circ$}. All other higher-order harmonics are evanescent. Efficient anomalous deflection into the $+\theta^{\rm r}$ direction is achieved by maximizing conversion into the $+1$ Bloch harmonic through optimization of an additional coupling network. The wide deflection angle is hand-picked, as conventional phase-gradient reflectors typically exhibit degraded directional performance under such conditions due to undesired parasitic scattering that can reach up to 30\% of the incident power. The unit-cell spacing of the three-patch linear sub-array configuration is set as \mbox{$D_x/3 = D_y = 0.3547\lambda$} ($3.798$~mm at $28$~GHz), with an overall supercell footprint of $1.064\lambda \times 0.3547\lambda$. The geometric parameters of the periodic RIS are outlined in Table~\ref{tab:dimentions} (second column).

We aim the periodic RIS to convert a plane wave with an $E$-field amplitude $\mathbf{E}^{\rm i}_0$ incident at angle $\theta^{\rm i}$ to a single plane wave with an amplitude $\mathbf{E}^{\rm s}_{+1}$ propagating in the desired direction $+\theta^{\rm r}$. The total scattered $E$-field amplitude $\mathbf{E}^{\rm s}_{+1}$, for a set of load impedances $\mathbf{Z}^\text{L}$, is given by a combined contributions of zero-current scattering and port-current scattering terms~\cite{vuyyuru23AWPL}:
\begin{equation}
\mathbf{E}^{\rm s}_{+1}(\mathbf{Z}^{\rm L})=
\mathbf{E}^{\rm s}_{+1}(\mathbf{I}^\text{A}=0)
+\sum_{m=1}^M
\frac{k\eta I^{\rm A}_m(\mathbf{Z}^{\rm L}) \bh^\text{A}_m(\theta^{\rm r})}{2D_xD_yk_{z,+1}},
\label{eq:Esm}
\end{equation}
where $k=2\pi/\lambda$ is the free-space wavenumber, $\eta\approx 377~\Omega$ is the free-space intrinsic impedance, and $k_{z,+1}=k\cos\theta^{\rm r}$. The phasor $\mathbf{E}^{\rm s}_{+1}(\mathbf{I}^{\mathrm{A}}=0)$ is the scattered $E$-field amplitude for the $(+1)$-th propagating Bloch harmonic, computed in the receiving  regime with all $M$ open-circuited array element terminals (see the treatment in Fig.~\ref{fig4:MEle-MLoadPort}). The array element current vector is found from $\mathbf{I}^{\rm A} = -\mathbf{I}^\text{O}=-\left(\mathbf{Z}^{\rm A}+\mathbf{Z}^{\rm O}\right)^{-1}\mathbf{V}^{\rm A}_\text{oc}$, where $\mathbf{V}^{\rm A}_\text{oc}$ represents open-circuit voltages obtained from the Rx simulation. The matrix $\mathbf{Z}^\text{O}$ depends on $\mathbf{Z}^\text{L}$ via Eq.~\eqref{eq:ZO}.

The global algebraic scattering synthesis approach treats load impedances $Z^{\rm L}_n$ $(n=1,\ldots,N)$ as optimization variables, avoiding the heavy computational cost of brute-force supercell-level optimization. Moreover, the rapid algebraic nature of optimization extends to any arbitrary large, finite apertures, mitigating the high computation cost of optimization associated with large, finite-sized reflectors. This design framework accounts for all EM mutual coupling among elements, reradiated and structural scattering, and possible non-ideal parasitic interactions, achieving EM-consistent RIS designs through a numerically efficient process.

The BD-RIS configuration utilizing the cascaded load network of Sec.~\ref{sec:methodology} is compared with the conventional Diagonal RIS (D-RIS) optimization with isolated loads directly connected to array element terminals. The D-RIS implementation serves as the baseline for comparison with the BD-RIS design. For carrying out the D-RIS optimization, preliminary full-wave simulations of the 3-element $(M=3)$ supercell are conducted using the CST Microwave Studio simulator to prepare an algebraic non-local optimization. Rx simulations with plane-wave illuminations from $\theta^i=0\degree,\theta^{\rm r}$ find $\mathbf{V}^\text{A}_\text{oc}$, $\bar{\bh}^\text{A}(\theta^{\rm r})$. From the Tx simulation we can find  $\mathbf{Z}^\text{A}$. The objective function maximizes the reflected power efficiency of the desired Bloch harmonic. For periodic structures, the efficiency of each Bloch harmonic is defined as the ratio of the reflected Bloch harmonic amplitude to that of the incident wave. In the D-RIS, the three patch ports are terminated in individual reactive loads that have no coupling between them. The optimization is implemented in MATLAB using the \verb|fmincon| function to find optimal loads for maximizing the efficiency. Details of the D-RIS optimization for periodic anomalous reflectors are available in \cite{vuyyuru23AWPL}.

For the BD-RIS with a cascaded load network, an equivalent $M$-port load network treatment is employed, as outlined in Sec.~\ref{subsec:MEle-MLoadPort}. The EM parameters extracted for D-RIS---$\mathbf{V}^\text{A}_\text{oc}$, $\bar{\bh}^\text{A}(\theta^{\rm r})$, and $\mathbf{Z}^\text{A}$---are also utilized for scattering synthesis of BD-RIS. An eight-port, non-radiating feed network is modeled at the circuit level in Keysight ADS, as illustrated in Fig.~\ref{fig:RISPeriodResults}(b), to extract the overall $Z$-parameter matrix, $\mathbf{Z}^{\rm F}$. Here, ports 1--3 $(M=3)$ are the output ports connected to the patch elements, and ports 4--8 $(N=5)$ are the input ports connected to impedance loads. The ADS model comprises microstrip transmission-line sections and cross junctions, interconnected to form a distributed $8$-port network terminated in $50~\Omega$ ports. The corresponding physical layout generated from the equivalent transmission-line model is depicted in Fig.~\ref{fig:RISPeriodResults}(c).  The microstrip segments capture guided-wave propagation and phase delay. The equivalent load-network impedance matrix, $\mathbf{Z}^{\rm O}$, as a function of the reactive load parameters $\mathbf{Z}^\text{L}$ via Eq.~\eqref{eq:ZO} is computed. Finally, a non-local optimization is carried out for maximizing the reflection efficiency using the total scattered $E$-field in Eq.~\eqref{eq:Esm}. As can be see in Fig.~\ref{fig:RISPeriodResults}(b), the load matrix $\mathbf{Z}^{\rm L}$, comprises five independent diagonal loads, with three (ports 4--6) connecting the antenna elements to ground and two (ports 7--8) interconnecting the neighboring antenna terminals, all via microstrip segments.

Figure~\ref{fig:RISPeriodResults}(d) compares the reflected power efficiencies of the conventional phase-gradient D-RIS design, the non-local optimized D-RIS, and the non-local optimized BD-RIS against the theoretical efficiency bound of an LPA metasurface (i.e., a reactive surface showing a linear reflection phase profile)~\cite{asadchy2017eliminating}. The phase-gradient design exhibits reduced efficiencies at extreme angles, and it underperforms relative to the LPA-based bound. In contrast, the non-local D-RIS optimization having a diagonal load matrix significantly improves the efficiency to 85\%, yet still below the ideal 100\% power efficiency limit for infinite reflectors. As was shown in~\cite{vuyyuru23AWPL}, increasing the number of elements per wavelength can attain the optimal efficiency for D-RISs, though at the cost of issues associated with array densification. Finally, the BD-RIS utilizing a non-locally optimized coupled load network achieves the nominally perfect reflection efficiency, without deep subwavelength structuring of the array. The developed BD-RIS optimization framework accounts for inter-element couplings via both EM field interactions above the ground plane and guided-wave coupling within the underlying load network.

\begin{figure}[ht]
 \centering
\includegraphics[width=3.45in,height = 3.15in]{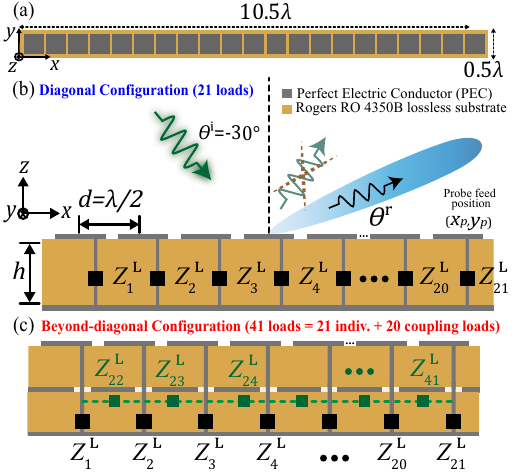}
\caption{(a) Top view of the 21-element linear array. (b) Side view with a diagonal load configuration comprising 21 $\lambda/2$-spaced elements, illustrating anomalous reflection from $\theta^{\mathrm{i}} = -30^\circ$ to $\theta^{\mathrm{r}}$. (c) Side view with a beyond-diagonal load configuration, where 20 additional adjacent coupling loads are introduced, providing 41 optimization parameters in the load $Z$-matrix.} 
\label{fig:RISlineararray}
\end{figure}

\subsection{Aperiodically loaded finite linear array} \label{subsec:finiteRIS}

An anomalous reflector design based on a finite one-dimensional linear array is implemented using $21$ $\lambda/2$-spaced microstrip patch elements along the $x$-axis, as illustrated in Fig.~\ref{fig:RISlineararray}(a). The RIS array spans an aperture of $10.5\lambda\times0.5\lambda$ in the $xy$-plane at $28$~GHz. The fixed periodic RIS structure described in Section~\ref{subsec:PeriodicRIS} features spatially periodic loading and geometry, which supports beam scanning only toward a discrete set of deflection angles, dictated by Bloch harmonics for the predetermined incident and the spatial period. This periodicity assumption forbids continuous angular scanning and neglects truncation effects, which is a limitation when modeling large finite-sized structures. Here, a finite RIS configuration is considered, and its geometric parameters are summarized in Table~\ref{tab:dimentions} (third column). The RIS is modeled for TE-polarized incidence at $\theta^{\rm i} = -30^\circ$ and a wide-angle anomalous reflection into multiple reconfigurable desired directions spanning $\theta^{\rm r} = 55^\circ$ to $70^\circ$ in $5^\circ$ increments. This wide-angle example illustrates the advantages of the proposed technique over the conventional reflectarray designs with a typical $\lambda/2$ element spacing, where linear reflection phase-gradient synthesis encounters challenges in achieving high efficiencies for such large anomalous deflections.

\begin{figure*}[t]
\centering
\includegraphics[width=7in,height = 4in]{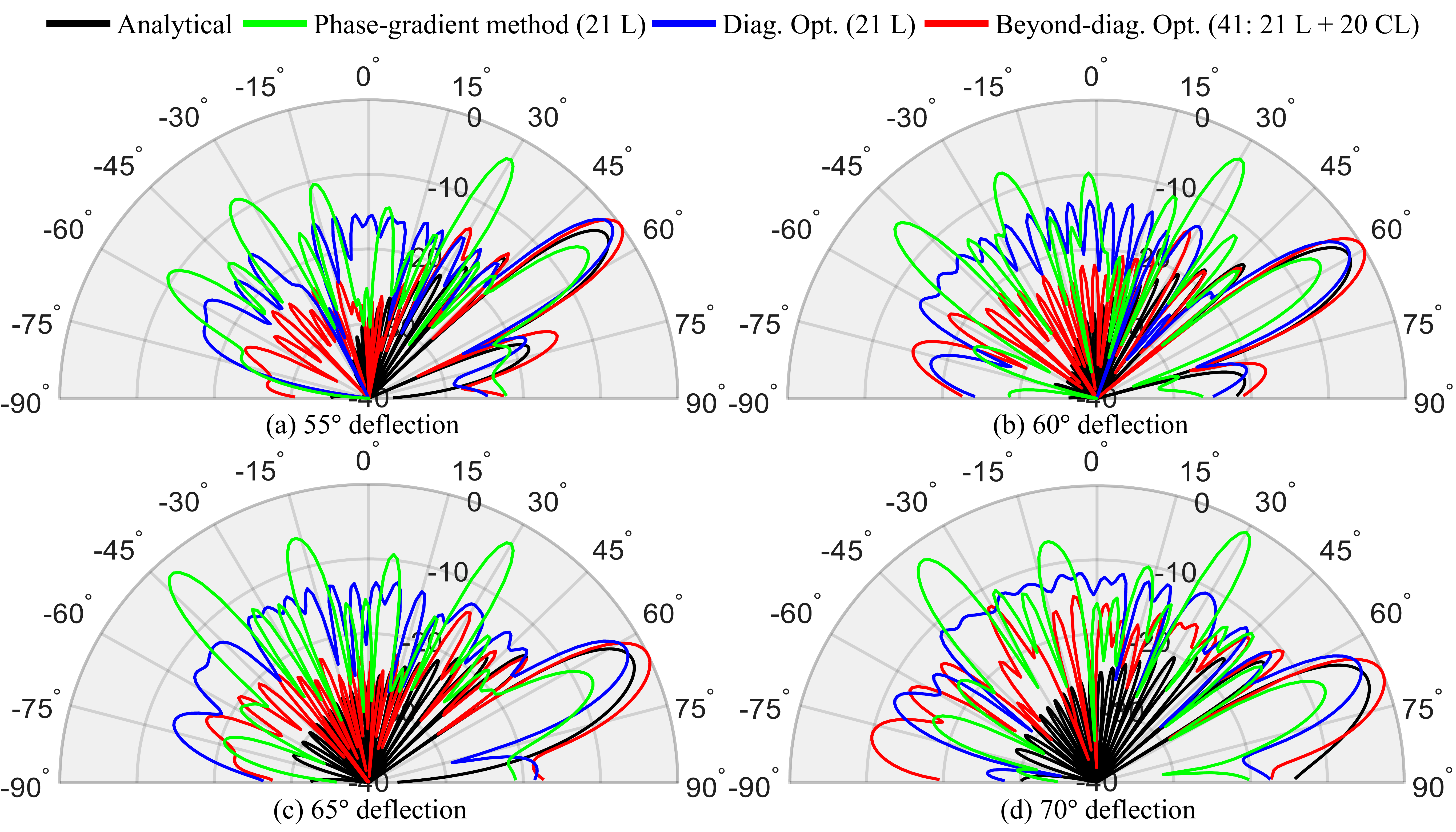}
\caption{Illustration of 1-D ($\phi = 0$) scattering cross section (SCS) patterns for (a) $55^\circ$, (b) $60^\circ$, (c) $65^\circ$, (d) $70^\circ$ designs of the linear anomalous reflector with a $\lambda/2$ element spacing in the $xz$-plane, expressed in dB after normalization. ``Analytical'' (solid black curve) denotes the reference 1-D SCS obtained from the physical optics approximation, modeling an ideal 100\%-efficient anomalous reflection from a hypothetical metasurface of the same size~\cite{MacroscopicARM2021}. These figures compare the conventional phase-gradient method (solid green curve), D-RIS synthesis (solid blue curve), and BD-RIS optimization (solid red curve). L: individual loads; CL: coupling loads.}
\label{fig:RISlineararrayresults}
\end{figure*}

Let the RIS be excited by a vertically polarized incident plane wave at angle $\theta^{\mathrm{i}}$, and the desired scattering direction is $\theta^{\mathrm{r}}$. The total scattered $E$-field, $\mathbf{E}^{\rm s}(\mathbf{Z}^{\rm L})$, in the far zone scattered by an $M$-element finite RIS in the predetermined direction $\theta^{\mathrm{r}}$ is given by the sum of zero-current and port-current scattering contributions~\cite{vuyyuru25OJAP}:
\begin{equation}\label{eq:radiated_E_field}
\mathbf{E}^{\rm s}(\mathbf{Z}^{\rm L}) = 
\mathbf{E}^{\rm s}(\mathbf{I}^{\mathrm{A}}=0)
+\sum_{m=1}^M 
\frac{jk\eta}{4\pi}
I^{\rm A}_{m}(\mathbf{Z}^{\rm L})
\mathbf{h}^{\rm A}_m(\theta^{\rm r})
\frac{e^{-jkr}}{r},
\end{equation}
where $\mathbf{h}^{\rm A}_m$ and $I^{\rm A}_{m}$ are the vector effective height and the port current of the $m$-th patch, respectively. The port current vector is expressed as $\mathbf{I}^{\rm A} =-\mathbf{I}^\text{O}=-\left(\mathbf{Z}^{\rm A}+\mathbf{Z}^{\rm O}\right)^{-1}\mathbf{V}^{\rm A}_\text{oc}$, where $\mathbf{V}^{\rm A}_\text{oc}$ and $\mathbf{Z}^{\rm A}$ denote open-circuit voltages and input impedance matrix of the patch array, respectively. The scattered $E$-field with open-circuit loads is denoted as $\mathbf{E}^{\rm s}(\mathbf{I}^{\mathrm{A}}=0)$. This formulation accounts for edge effects and inter-element mutual coupling among all radiating patches, independent of the incident field and loading conditions. A multi-objective non-linear algebraic synthesis method is employed to synthesize the scattering characteristics of the passively loaded patch array by parameterizing the lumped reactive load impedances, $Z^\text{L}_n$ $(n=1,\ldots,41)$, as optimization variables, thereby avoiding repeated computationally intensive full-wave EM simulations at every iteration in the optimization loop.

Similar to Sec.~\ref{subsec:PeriodicRIS}, three configurations are evaluated to assess the efficacy of the proposed non-local cascaded load network. The first corresponds to the conventional phase-gradient D-RIS employing reactive loading to realize a linear reflection phase profile along the array. The second considers non-local optimization of a D-RIS with $21$ individual loads ($M=21$), as depicted in Fig.~\ref{fig:RISlineararray}(b). The third extends this approach to a BD-RIS using a cascaded load network with $41$ loads ($N=41$). To begin, preliminary full-wave EM simulations of the $M$-element RIS are conducted to acquire $\mathbf{Z}^{\rm A}$, $\mathbf{V}^{\rm A}_\text{oc}$, and $\bar{\mathbf{h}}^{\rm A}$ for D-RIS optimizations. Additionally for the BD-RIS, the 62-port non-radiating feed network is modeled at the circuit level to obtain $\mathbf{Z}^{\rm F}$. During the BD-RIS optimization, the equivalent $M$-port load network $Z$-parameter $\mathbf{Z}^{\rm O}$ is computed using~\eqref{eq:ZO} for each candidate $\mathbf{Z}^\text{L}$.

\begin{table}[t]
\renewcommand{\arraystretch}{1.25}
    \begin{center}
    \caption{Comparison of the SLL and optimized efficiency between diagonal and beyond-diagonal load configurations.}
    \label{tab:eff}
    \footnotesize
    \begin{tabular}{|c|c|c|c|c|c|c|} \hline
    \multirow{2}{*}{\shortstack{Desired \\ angle $(\theta^{\rm r})$}} & \multicolumn{2}{c|}{Refl. (21 L)} & \multicolumn{2}{c|}{Diag. Opt. (21 L)} & \multicolumn{2}{c|}{B-diag. Opt. (41 L)} \\ \cline{2-7}
    & SLL & Eff.  & SLL & Eff. & SLL & Eff.\\ \hline
    55$^\circ$ & +1.4 & 47.8 & -13.8 & 121.5 & -15.2 & 169.2 \\ \hline
    60$^\circ$ & +3.8 & 34.83 & -11.3 & 100 & -14.5 & 163.7 \\ \hline
    65$^\circ$ & +5.8 & 26 & -9 & 76 & -13.7 & 163.7 \\ \hline
    70$^\circ$ & +7.5 & 21.8 & -6.5 & 60.8 & -10 & 162.3 \\ \hline
    \end{tabular}
    \end{center}
\end{table}

For reference, the analytically predicted scattering cross section (SCS) pattern is evaluated, assuming ideal anomalous reflection with a 100\% efficiency from a perfectly reflecting impedance boundary with continuous surface current distribution over the same aperture size as the three RISs. A physical-optics-based analytical model is adopted, appropriate for electrically large reflectors relative to the wavelength. For a rectangular aperture of dimensions $D_x\times D_y$  under a TE-polarized plane wave reflection in the $xz$-plane, the analytical expression for the scattered $E$-field is given as~\cite{MacroscopicARM2021}
\begin{align}
    \mathbf{E}^\text{s}_\text{ref} (\theta) &= \frac{j k e^{-jkr}}{4 \pi r} E_0 S\Bigg[\left( \cos\theta - \cos\theta^{{\rm i}} \right) \, {\rm sinc}(ka_{\rm ef}) \notag \\ 
    & \hspace{1cm} + r_n\left( \cos\theta + \cos\theta^{\rm r}\right) {\rm sinc}(ka_{{\rm ef}n})\Bigg],
\label{eq:scatt_ref}
\end{align}
where $a_{\rm ef} = D_x(\sin\theta - \sin\theta^{\rm i})/2$ and $a_{{\rm ef}n} = D_x(\sin\theta - \sin\theta^{\rm r})/2$ characterize the effective sizes of the metasurface for the incident and reflected waves, respectively. Here, sinc denotes the sinc function [$\mathrm{sinc}(x)=(\sin x)/x$], and $S$ represents the aperture area. The macroscopic reflection coefficient $r_n$ is defined as the ratio of the perfectly reflected-field amplitude in the desired direction to that of the incident wave. Power conservation in plane-wave reflection from an infinite boundary yields $r_n=\sqrt{\cos\theta^{\rm i}/\cos\theta^{\rm r}}$. For an ideal anomalous reflector, only the $r_n$ associated with the desired deflection direction is nonzero. Using \eqref{eq:scatt_ref} as the analytical benchmark, the efficiencies as well as the far-zone patterns of the three designs are compared. The efficiency $\zeta$ is defined as the ratio of the far-zone  power density of the field reflected by the optimized RIS to that of the ideal anomalous reflector. The definition reads
\begin{equation}
\zeta = 
\frac{\vert \mathbf{E}^{\rm s}(\mathbf{Z}^{\rm L},\theta^{\rm r}) \vert ^2} 
{\vert \mathbf{E}^\text{s}_{\text{ref}}(\theta^{\rm r}) \vert ^2},
\label{eq:finiteefficiency}
\end{equation}
where $\mathbf{E}^{\rm s}(\mathbf{Z}^{\rm L},\theta^{\rm r})$ denotes the computed scattered field in the $\theta^{\rm r}$ direction for the array loaded with $\mathbf{Z}^{\rm L}$.

Figure~\ref{fig:RISlineararrayresults} plots the SCS patterns of the linear anomalous reflector with a $\lambda/2$ element spacing for all three cases, along with the analytical prediction using Eq.~\eqref{eq:scatt_ref}. Each plot is normalized by the maximum value of the four SCS curves, which is always achieved by the BD-RIS design. The efficiency and side-lobe level (SLL) are computed and summarized in Table~\ref{tab:eff}, where the SLL is defined as the ratio of the bistatic SCS of the strongest unwanted minor scattering lobe to that of the desired anomalous scattering lobe. As anticipated, the conventional phase-gradient design exhibits a pronounced efficiency degradation and high SLLs for large anomalous reflection angles with elevated undesired minor scattering lobes, especially in the specular direction. The D-RIS optimization enhances the efficiency and suppresses SLLs, demonstrating a dominant scattering beam into the desired anomalous direction while minimizing the specular lobe and other undesired reflections in Fig.~\ref{fig:RISlineararrayresults}. It exhibits a superdirective behavior at the $55^\circ$ deflection angle, where ``superdirective'' refers to solutions with the efficiency $\zeta$ of Eq.~\eqref{eq:finiteefficiency} that exceed unity. However, for extreme steering angles ($65^\circ$ and $70^\circ$), the efficiency drops below the ideal 100\% level. Moreover, the scattering maximum is not aligned precisely with the desired angle, leading to a beam pointing error (BPE) that becomes more pronounced at large deflection angles. This deviation arises from the relatively coarse half-wavelength inter-element spacing and the patch element radiation pattern, which significantly weakens at grazing angles.

In contrast, the proposed BD-RIS optimization with a cascaded load network achieves highly directive anomalous reflection with substantially improved efficiencies, suppressed minor lobes, and diminished BPEs. The beyond-diagonal configuration with 41 loads achieves superdirective efficiencies across all desired angles by fully suppressing the specular and undesired reflections, as reported in Table~\ref{tab:eff}. Superdirectivity generally relies on controlling evanescent fields through a non-local subwavelength-loaded configuration, whereas sparse configurations with one element per $\lambda/2$ inherently limit their manipulation. However, these results show that superdirectivity can be attained using conventional arrays with a $\lambda/2$ spacing, without the need to control the subwavelength current distribution on the reflecting RIS. The superdirective behavior results from controlled inter-element interactions via global non-local optimization of reactive loads in the cascaded network, suppressing undesired parasitic scattering through additional degrees of freedom of a virtual array. Overall, the non-local cascaded-load optimization demonstrates substantial performance enhancement without requiring array densification.

\section{Conclusion}\label{sec:conclusion}

This study demonstrates and validates an EM-consistent co-simulation pathway for realizing non-local BD-RIS designs with optimized scattering characteristics without relying on subwavelength structural complexity of the array. By generalizing the conventional diagonal load impedance optimization with a non-diagonal form implemented through a non-local cascaded coupling load network, the performance of the reflectarray and RIS enhances with increased degrees of freedom. The proposed technique incorporates both EM near-field coupling through antenna characterization and additional tunable guided-wave coupling beneath the ground plane via circuit-level modeling.

A computationally efficient RIS synthesis framework based on fixed-period linear and periodic supercell arrays with realistic patch elements has been presented for anomalous reflection. Beyond-diagonal configurations consistently outperform diagonal designs. In periodic supercell arrays supporting a finite set of Bloch harmonics, a perfect efficiency is achieved for a wide-angle deflection using an element density with which D-RIS designs are sub-optimal. For finite arrays, a cascaded load network with a half-wavelength inter-element spacing significantly surpasses the conventional phase-gradient and optimized D-RIS approaches, particularly at extreme deflection angles. Beyond the specific designs illustrated here, this approach is broadly generalizable to arbitrary structural geometries and incident wave polarizations.

By decoupling the non-local functionality from subwavelength spatial structuring requirements, the approach reduces the array element density and provides strong control of nonlocal electromagnetic response of the RIS. In general, a trade-off inherently arises between implementation complexity and performance; however, the results demonstrate that BD-RIS configurations attain high efficiency even with conventional half-wavelength inter-element spacing. Future work will focus on demonstrating a wide range of reconfigurable advanced RIS functionalities using the generalized beyond-diagonal cascaded coupling network synthesis.

%\appendices
%\section{Non-diagonal load matrix calculation from circuit components using Kirchhoff’s laws}\label{sec:appendix_A}

%\section{Toward Realistic Modeling: Numerical Analysis Accounting for Material Losses}\label{sec:losses}
 
\ifCLASSOPTIONcaptionsoff
  \newpage
\fi

%\clearpage
%\newpage
\balance
\bibliographystyle{IEEEtran}
\bibliography{IEEEabrv,Bibliography}

\end{document}